\documentclass[11pt,a4paper]{article}
\usepackage[utf8]{inputenc}
\usepackage[english]{babel}
\usepackage{amsmath}
\usepackage{amsfonts}
\usepackage{amssymb}
\usepackage{lmodern}
\usepackage{booktabs}
\usepackage{pdfpages}

\usepackage[utf8]{inputenc}

\newcommand{\cm}{$c$}
  
\newcommand{\reffigure}[1]{Figure~\ref{#1}}
\newcommand{\reftable}[1]{Table~\ref{#1}}
\newcommand{\refsection}[1]{Section~\ref{#1}}
\newcommand{\refequation}[1]{Equation~(\ref{#1})}

\begin{document}

\title{The impact of partially missing communities~on the reliability of centrality measures}
\author{Christoph Martin}
\date{%
Leuphana University of L{\"u}neburg \\
 Universitätsallee 1, 
         21335 L{\"u}ne­burg, Germany \\
\texttt{cmartin@uni.leuphana.de} \\
\vspace{0.5cm}
    September 20, 2017
}

\maketitle

\begin{abstract}

Network data is usually not error-free, and the absence of some nodes is a very common type of measurement error. Studies have shown that the reliability of centrality measures  is severely affected by missing nodes. This paper investigates the reliability of centrality measures when missing nodes are likely to belong to the same community. We study the behavior of five commonly used centrality measures in uniform and scale-free networks in various error scenarios. 
We find that centrality measures are generally more reliable when missing nodes are likely to belong to the same community
than in cases in which nodes are missing uniformly at random.
In scale-free networks, the betweenness centrality becomes, however, less reliable when missing nodes are more likely to belong to the same community. Moreover, centrality measures in scale-free networks are more reliable in networks with stronger community structure. In contrast, we do not observe this effect for uniform networks.
Our observations suggest that the impact of missing nodes on the reliability of centrality measures might not be as severe as the literature suggests.
\end{abstract}

\section{Introduction}
\label{sec:intro}
Centrality measures are commonly used in network analysis. Network data is, however, rarely error-free. Some parts of a network are often not recorded correctly. For example, nodes could be missing due to the non-response effect or the boundary specification problem \cite{Kossinets2006}. 
Some users in social networking services may have restrictive privacy settings and thus their profiles are not accessible or the number of API requests might be limited \cite{Rezvanian2015}. In addition, this type of error does commonly occur when bio-logging is used to collect interaction data~\cite{silk2017}.

Studies have found that the reliability of centrality measures is often severely compromised by missing nodes \cite{Smith2017, Smith2013, Boldi2013, Wang2012, Frantz2009a, Kim2007, Borgatti2006a, Kossinets2006, Costenbader2003, Bolland1988}.
These studies agree that higher levels of error lead to lower reliability.  The exact extent of the impact of missing nodes on the reliability does, however, depend on a variety of factors. Analyzing Erdős–Rényi networks, Borgatti et al. \cite{Borgatti2006a} observed that centrality measures  behaved  similarly. 
Considering different types of empirical networks and random graphs, studies found that the reliability of centrality measures strongly depends on the type of network. 
For example, Smith et al. \cite{Smith2017} found that centrality measures are more reliable in larger, more centralized networks. Closeness centrality was been reported to be more reliable than betweenness and degree centrality~\cite{Kim2007}. Moreover, Boldi et al. \cite{Boldi2013} observed that social networks are more robust to missing nodes than web graphs.
Despite their important findings, previous studies have mostly focused on the case where nodes are missing uniformly at random. A notable exception is \cite{Smith2017}. In this study, the authors investigated the effect of missing nodes in cases in which the probability that a node is missing depends on their centrality. They found that the reliability is worse when more central nodes are missing.

Since nodes in a network are interconnected, it seems obvious that the 
behavior of nodes in a community will, at least to some extent, determine whether other nodes from that community can be observed or not \cite{Smith2017}.
For example,  some groups within a network may have concerns about the collection of their data and therefore collectively refuse to participate in a survey or adopt strict policies regarding the use of their data in social networking services. In animal research, subgroups of a population may be able to avoid being trapped and tagged and are therefore missing in the resulting network.
Despite the multitude of possible scenarios in which this type of measurement error may occur, it has not been considered in previous research.

In this study, we investigate how reliable centrality measures are when missing nodes are likely to belong to the same community, i.e., the probability that a node is missing depends on which other nodes are missing.\footnote{
Despite the similar wording, this work does not address the reliability or robustness of networks. For an extensive overview about the robustness of networks see~\cite{Havlin2010, Barabasi2016}.}
In particular, we examine whether this type of measurement error has a stronger or smaller impact on the reliability than the purely random absence of nodes. 
We use two random graph models to answer this question for uniform and scale-free networks. These models enable us to analyze the influence of the community structure on the reliability. 

Our results suggest that centrality measures are more reliable when missing nodes are likely to belong to the same community than in cases in which nodes are missing uniformly at random.
In scale-free networks, however, the betweenness centrality becomes less reliable when missing nodes are more likely to belong to the same community. Moreover, in scale-free networks,  centrality measures are more reliable in networks with stronger community structure. In contrast, we do not observe this effect for uniform networks. 
In addition to presenting these findings, we introduce a novel approach which we will refer to as "community bias". This approach allows us to simulate different levels of measurement error and enables us to study their impact on the reliability of centrality measures.

\section{Methods \& experimental setup}
\label{sec:methods}
We denote an undirected, unweighted graph by $G$ and the vertex set of a graph $G$ by $V(G)$.
In all graphs that we consider in this study, every node belongs to some community. We denote the nodes that belong to community $j$ in graph $G$ by $V_j(G)$.
We use the terms graph and network interchangeably. A centrality measure \cm\ is a real-valued function that assigns centrality values to all nodes in a 
graph and is invariant to structure-preserving mappings, i.e., centrality values depend solely on the structure of a graph. External information (e.g., node or edge attributes) have no influence on the centrality values \cite{Koschutzki2005}.
We denote the centrality value for node $u \in V(G)$ by $c_G(u)$ and the centrality values for all 
nodes in G~$(u_1, u_2, \dots, u_n)$ by the vector $c(G) := (c(u_1), \dots, c(u_n))$. 

The following centrality measures are used in this study: closeness centrality~\cite{Freeman1978}, betweenness centrality \cite{Freeman1977}, degree centrality, eigenvector centrality \cite{Bonacich1987}, and the PageRank \cite{Brin1998}.

There are multiple definitions of communities in networks and, depending on the context, some are more appropriate than others. In this study, a community is a subgraph where each of its vertices is more strongly attached to vertices in that subgraph than to vertices in any other subgraph \cite{Hu2008, Fortunato2016}.
Hence, the fraction of edges that a node has to other nodes which are not part of its community (compared to the total number of edges that are connected to this node) is an indicator of the strength of the community structure ("community strength") in a network. 
The lower this ratio, the stronger the community structure. We can quantify the strength of the community structure by calculating the modularity of a graph with respect to a mapping which maps the nodes to communities \cite{Newman2004}.

Some community definitions allow communities to overlap. We focus on non-overlapping communities. For a more detailed discussion of the definition of communities in networks see~\cite{Wasserman1994, Boccaletti2006, Fortunato2010}.

\subsection{Data}
\label{sec:data}
To investigate the effect of community structure on the reliability of centrality measures, we use two random graph models. There are two main reasons to use synthetic graphs. When using random graphs models, we know the ground-truth mapping from nodes to communities, i.e., we know which nodes belong to the same community. In contrast, for real-world networks we might not know if there are communities at all \cite{Fortunato2007}.  Moreover, the random graph models enable us to vary the strength of the community structure (as described above) and thus gives us the opportunity to study the effect of the community strength on the reliability of centrality measures explicitly.

In the clustered random graph (CRG) model,  $n$ nodes are partitioned into $k$ sets. Nodes in the same set belong to the same community. With probability $p_{intra}$, edges are created between nodes in the same community. Edges between nodes that are not in the same community are created with probability $p_{inter}$. This model was originally introduced by \cite{Girvan2002} to benchmark community detection algorithms. Since the number of edges from a node to other nodes in the same community and the number of edges from a node to nodes in other communities both follow a binomial distribution (with different parameters though), this model is conceptually close to Erdős–Rényi graphs \cite{Erdos1959}.

We use two configurations of the CRG model, one with weaker community structure (CRG\textsubscript{weak}) and one with stronger community structure (CRG\textsubscript{strong}). In both configurations, we set $n=1000$ and $k=25$.
For the CRG\textsubscript{weak} configuration, we set $p_{intra}$ to $0.1$  and $p_{inter}$ to $0.01$.
For the CRG\textsubscript{strong} configuration, we set
$p_{intra}$ to $0.2$ and $p_{inter}$ to $0.005$.

The second model is the Lancichinetti–Fortunato–Radicchi (LFR) model as described in \cite{Lancichinetti2008, Staudt2017}. According to this model, the distribution of the node degrees and distribution of the community sizes both follow a power-law distribution.
The degree distribution and the community size distribution in empirical networks can often be described by a power-law distribution \cite{Leskovec2008, Clauset2009}.
Hence, graphs generated by this models share various characteristics with real-world networks. 

For the degree distribution, we use an average degree of 10, a maximum degree of 50, and an exponent of $-2$. For the community size distribution, we use minimum community size of 5, maximum community size of 100, and an exponent of $-2$.
The mixing parameter $\mu$ determines the fraction of neighbors of each node that do not belong to the node's own community. 
Again, we use two configurations of the LFR model. One with $\mu = 0.8$ and thus a  weaker community structure  (LFR\textsubscript{weak}) and one with $\mu = 0.4$ and thus a stronger community structure (LFR\textsubscript{strong}). In addition, we use a third variation of this model for the second part of our experiments. Here we use the same parameters as described above, but vary the mixing parameter $\mu$ from $0.15$ to $0.95$ in steps of $0.05$. We denote these 17 configurations by LFR\textsubscript{varying}($\mu$). 

In both of these random graph models, the centrality measures are usually correlated. However, this is not problematic since centrality measures in real-world networks are also often correlated with each other \cite{valente2008correlated}. Various properties of graphs that are generated by the random graph configurations used in this paper are listed in \reftable{tab:randomgraphstatistics}.

%
\begin{table}
\caption{Statistics for graphs generated by the random graph configurations that are used in this paper. Numbers are mean values based on 100 realizations. Standard deviations are listed in parentheses. "Clustering" denotes the average clustering coefficient and "Communities" denotes the number of communities. 
}
\label{tab:randomgraphstatistics}       
%
%
\resizebox{\textwidth}{!}{

    \begin{centering}
\begin{tabular}{lrrrrrr}
\hline\noalign{\smallskip}
       &      Nodes &        Edges &   Diameter & Communities & Clustering &     Modularity \\
\hline
 CRG\textsubscript{strong} &   1000  &   6380  (86) &  5.0 (0.1) &    25 (0.0) &   0.08 (0.003) &  0.581 (0.006) \\
   CRG\textsubscript{weak} &   1000  &   6798 (78) &  4.9 (0.3) &    25 (0.0) &   0.02 (0.001) &  0.253 (0.006) \\
 LFR\textsubscript{strong} &   1000  &   5028 (129) &  6.0 (0.0) &    69 (7.9) &   0.18 (0.017) &  0.534 (0.007) \\
   LFR\textsubscript{weak} &   1000  &   5022 (126) &  5.0 (0.2) &    68 (8.6) &   0.03 (0.002) &  0.149 (0.007) \\
\hline
\end{tabular}

\end{centering}
}
\end{table}

\subsection{Quantifying measurement errors and reliability}

\subsubsection{Modeling measurement errors}

\label{sec:errormechanism}

As discussed in the introduction, in a variety of scenarios, it is reasonable to assume that missing nodes are not independent of each other. In fact, missing nodes might belong to the same community and thus the absence of nodes is "biased" towards communities. Here we describe a novel approach to model this type of measurement error.

To simulate that $\left \lceil \alpha \cdot | V(G)| \right \rceil$ nodes are missing from a graph $G$, we create a copy of $G$ that we denote by $G'$ and proceed as follows:
\begin{enumerate}
\item
 First, we choose a community from which one node will be removed. We denote this community by $j$. We can enumerate the communities since the random graph models provide us the mapping from the nodes to the communities.

Let $P(j)$ be the probability that community $j$ will be selected. Then

\begin{equation}
\label{eq:propbias}
P(j) \propto \left [ 1 + missing(j) \right ]^{\lambda}
\end{equation}

if there are still nodes in the graph that belong to community $j$. Here, $missing(j)$ denotes the number of nodes that belong to community $j$ and have already been removed from the graph $G'$ and $\lambda$ is a non-negative real number which determines the strength of the "community bias". If all nodes of community $j$ have already been removed from the graph, $P(j) = 0$.
\item
Next, we randomly choose a node from $V_j(G)$ that has not yet been removed from $G'$ and remove it from $G'$
\item
We repeat this process until $\left \lceil \alpha \cdot | V(G)| \right \rceil$ nodes have been removed from $G'$.
\end{enumerate}

This procedure has two parameters. The intensity of the simulated measurement error is controlled by $\alpha$, the fraction of nodes that are removed from the graph.
The extent of the bias of missing nodes to belong to the same community ("community bias") is controlled by $\lambda$.
If $\lambda = 0$, then there is no community bias and all nodes have the same probability to be removed from the graph, independently of already missing nodes.
For $\lambda > 0$, nodes are more likely to be removed from communities where nodes have already been removed.
For large values of $\lambda$, the community that gets chosen in the first iteration usually gets chosen again and again until all nodes from that community are removed from the graph. In this case, entire communities are essentially removed successively.

\subsubsection{Quantifying the reliability}
\label{sec:defnreliability}
Network data is usually affected by measurement errors, as discussed in the introduction (e.g., some actors are missing).
Hence, we seek to reveal  the reliability of centrality values that are calculated based on the erroneous network data. To quantify this reliability of centrality measures, we use the Kendall  tau-b rank correlation coefficient $\tau$ \cite{Kendall1945}.
Rank correlations are commonly used to evaluate the ramifications of network modifications on centrality measures because researchers are often interested in the ranking of nodes derived from centrality measures rather than in the actual centrality values \cite{Kim2007, Wang2012, Lee2015}.

Let $G$ be the "error-free" graph, $G'$ an erroneous version of $G$ which is affected by some type of measurement error,  and $c$ a centrality measure. We define the reliability of the centrality measure $c$ with respect to $G$, $G'$, and the type of measurement error as $\tau (c(G), c(G'))$. Similar to existing studies, we only consider entries in $c(G)$ and $c(G')$ which correspond to nodes that do exist in $G$ and $G'$ \cite{Kim2007, Wang2012}. Moreover, we only consider nodes that are in the largest connected component of the particular  graph. (We observed, however, that almost all graphs in our experiments were connected.)
For reasons of brevity, we only write $\tau$ for the reliability of a centrality measure $c$ when $G$, $G'$, and $c$ are apparent from the context.

\subsection{Experimental setup}

For all random graph configurations that are described in \refsection{sec:data}, we study the impact of erroneous data collection on the reliability of centrality measures as follows:
\begin{enumerate}
\item
Generate a graph according to the random graph configuration (e.g., LFR\textsubscript{weak}) and denote it by $G$.
\item
Apply the remove node procedure (\refsection{sec:errormechanism}) with parameters $\alpha$ and $\lambda$ to $G$ and denote the resulting modified graph by $G'$.
\item
Finally, calculate the reliability of the centrality measures $\tau (c(G), c(G'))$ as described above (\refsection{sec:defnreliability}).
\end{enumerate}

\noindent
For our experiments, we use the following parameters:
As centrality measures $c$ we use betweenness, closeness, degree, eigenvector centrality, and PageRank.
As the fraction of nodes that are removed from the graph, we use values of $\alpha$ ranging from $0.025$ to $0.5$ in steps of $0.025$.
To control the extent of the community bias (the likelihood that  missing nodes belong to the same community), we use values of $\lambda$ ranging from $0$ to $3$ in steps of $0.5$.
For all combinations of these parameter values, we perform the experiment $100$~times.

The NetworkKit library (\cite{staudt_sazonovs_meyerhenke_2016}, v4.3) is used for graph generation and calculation of centrality measures. The NetworkX library (\cite{Hagberg2008}, v1.11) is used for various graph modifications.

\subsection{Statistical analysis}
\label{sec:modelDesc}
In addition to a visual inspection, we use two linear models  to investigate the relationship between the reliability of centrality measures and the error level, the community bias, and the strength of communities.

To analyze the results for the configurations CRG\textsubscript{weak}, CRG\textsubscript{strong}, LFR\textsubscript{weak}, and LFR\textsubscript{strong}, we use the following model:

\begin{equation}
\label{eq:modelFourGraphs}
\tau = \beta_0 + \beta_{1}^{i,j} \cdot\sqrt{\alpha} + \beta_{2}^{i,j} \cdot \sqrt{\alpha} \cdot \lambda + \epsilon 
\end{equation}

\noindent
With $i$ and $j$ as indices for the centrality measure and graph configuration, respectively.
This allows us to have different coefficients for each centrality measure and graph configuration. The error term is denoted by $\epsilon$.

To analyze the results of the experiments regarding the LFR\textsubscript{varying}($\mu$) models~(with $\mu$ ranging from $0.15$ to $0.95$ in steps of $0.05$), we use the following model:

\begin{equation}
\label{eq:modelVaryingGraphs}
\tau = \beta_0 + 
\beta_{1}^{i,j} \cdot \sqrt{\alpha} + 
\beta_{2}^{i,j} \cdot \sqrt{\alpha} \cdot \mu +
\beta_{3}^{i,j} \cdot \sqrt{\alpha} \cdot \lambda + 
\beta_{4}^{i,j} \cdot \sqrt{\alpha} \cdot \mu \cdot \lambda + \epsilon 
\end{equation}

\noindent
With $i$ and $j$ as indices for the centrality measure and graph configuration, respectively. The error term is denoted by $\epsilon$.  

We use the square root function to take into account observations from previous studies which have revealed a non-linear relationship between missing nodes and reliability \cite{Smith2013, Smith2017}. Moreover, our experiments have shown that these models provide a better fit to the data than models which do not use this transformation.

%
\begin{figure}
\includegraphics[width=\textwidth]{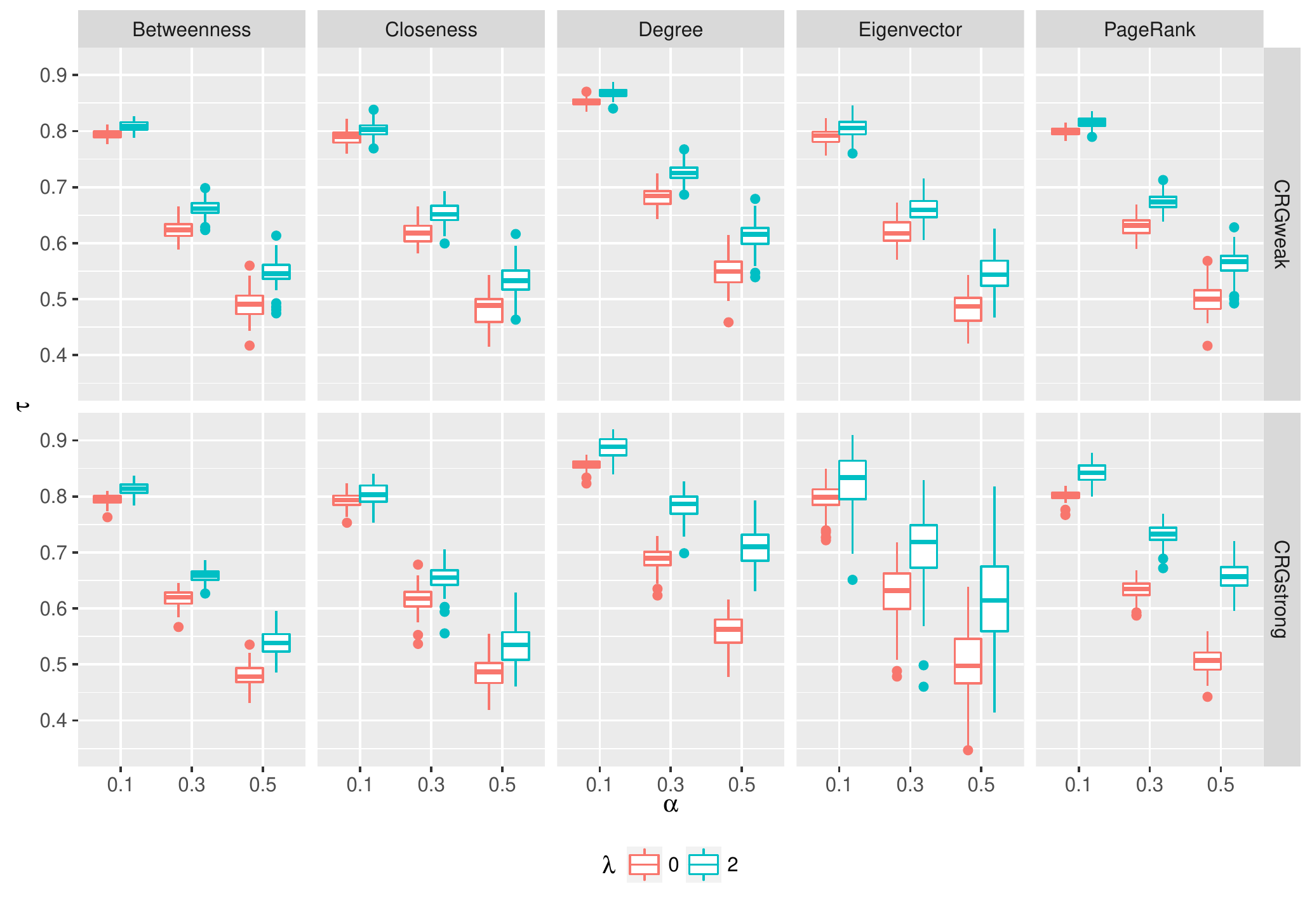}
%
%
\caption{
The figure shows results for the CRG models. For better visibility, the plot only contains results for $\lambda \in \{0,2\}$ and $\alpha \in \{0.1, 0.3, 0.5\}$. The bottom and top of the boxes indicate the first and third quartiles, respectively. The thick line within the box indicates the median.
}
\label{fig:boxplot}       
\end{figure}

\section{Results}
As outlined in the introduction, network data is often affected by measurement errors and in many cases, it is reasonable to assume that there is some dependency between the nodes that are missing. The goal of this study is to investigate how reliable centrality measures are when network data is incomplete and the missing nodes are likely to belong to the same community.

In general, our results suggest that centrality measures are more reliable when missing nodes are biased to belong to the same community. Moreover, for scale-free networks (LFR model) we observe that centrality measures are more reliable in networks with stronger community structure. However, we also observe that, in scale-free networks, the betweenness centrality becomes less reliable with increasing bias.

    \reffigure{fig:boxplot} illustrates the results for the graphs generated by the CRG models. For better visibility, the plot only contains results for $\lambda \in \{0,2\}$ and $\alpha \in \{0.1, 0.3, 0.5\}$. The bottom and top of the boxes indicate the first and third quartiles, respectively. The thick line within the box indicates the median.
As can be seen, even small levels of error result in a considerable drop (ranging from $0.1$  to  $0.2$) of the reliability. Moreover, all centrality measures are generally more reliable when the missing nodes belong to the same community ($\lambda = 2$) compared to uniform node missingness ($\lambda = 0$).
This effect is more noticeable in cases with stronger community structure (CRG\textsubscript{strong}).
We also notice that the variance of the reliability increases with increasing error level. It is particularly high for the eigenvector centrality and lowest for the degree centrality.

For a more detailed analysis of the relationship between the bias of missing nodes (controlled by $\lambda$) and the reliability, we use the model shown in \refequation{eq:modelFourGraphs} from \refsection{sec:modelDesc}.
(We also performed our analyses using more robust methods (i.e., weighted linear regression and quantile regression) and these results are consistent with the results reported in this section.)
The coefficient and standard error estimates for this model are listed in \reftable{tab:summaryFourGraphs}.
These results confirm our previous observation for the CRG configurations: higher community bias is related to higher reliability (interaction term~$\sqrt{\alpha} \cdot \lambda$).
For the LFR\textsubscript{strong} configurations, this effect only occurs for the degree centrality and the PageRank. For the closeness centrality, we observe the opposite effect. For the betweenness and eigenvector centrality, the coefficients are small and the effect is negligible.
The coefficients of the interaction term regarding the LFR\textsubscript{weak} model are significant but small, the effect is hardly noticeable.
Comparing the CRG and the LFR model, the effect of $\lambda$ on the reliability is usually stronger in graphs generated by one of the CRG models.

To analyze the impact of the community strength (controlled by $\mu$) on the reliability of centrality measures, we use the model shown in \refequation{eq:modelVaryingGraphs}.
The coefficient and standard error estimates for this model are listed in \reftable{tab:summaryLFRmu}.
The coefficients for  $\sqrt{\alpha}$ and  $\sqrt{\alpha} \cdot \mu$ are in good agreement with the results of the first model~(\refequation{eq:modelFourGraphs}).

All centrality measures except the betweenness centrality become more reliable with increasing strength of the community structure (indicated by $\sqrt{\alpha} \cdot \mu$). The contrary is true for the betweenness centrality.
The results also show  (indicated by $\sqrt{\alpha} \cdot \mu \cdot \lambda$) that, in case of betweenness, degree centrality and PageRank, a bias of missing nodes towards community amplifies the previously mentioned effect. The contrary is true for the closeness centrality, though the effect is small. In case of the eigenvector centrality, the effect is negligible.

%
\begin{table}
\caption{
Results for the model in \refequation{eq:modelFourGraphs}. Standard errors are listed in parentheses. \newline Intercept: 1.034 *** (1.9E-04), adjusted $R^2$:  $0.938$, p-value: $< 2.2E-16$
}
\label{tab:summaryFourGraphs}       
%
%
\resizebox{\textwidth}{!}{
\begin{tabular}{lrrrrrrrrr}
\hline\noalign{\smallskip}
            &  & \multicolumn{2}{l}{CRG\textsubscript{strong}} & \multicolumn{2}{l}{CRG\textsubscript{weak}} & \multicolumn{2}{l}{LFR\textsubscript{strong}} & \multicolumn{2}{l}{LFR\textsubscript{weak}} \\
            &  & $\sqrt{\alpha}$ & $\sqrt{\alpha} \cdot \lambda$ & $\sqrt{\alpha}$ & $\sqrt{\alpha} \cdot \lambda$ & $\sqrt{\alpha}$ & $\sqrt{\alpha} \cdot \lambda$ & $\sqrt{\alpha}$ & $\sqrt{\alpha} \cdot \lambda$ \\
\hline
Betweenness &&     -0.783 &                0.039 &     -0.766 &                0.034 &     -0.567 &                0.002 &     -0.580 &                0.005 \\
            &&        *** &                  *** &        *** &                  *** &        *** &                  *** &        *** &                  *** \\
            &&  (9.6E-04) &            (5.0E-04) &  (9.6E-04) &            (5.0E-04) &  (9.6E-04) &            (5.0E-04) &  (9.6E-04) &            (5.0E-04) \\
Closeness &&     -0.813 &                0.046 &     -0.782 &                0.032 &     -0.340 &               -0.031 &     -0.494 &               -0.008 \\
            &&        *** &                  *** &        *** &                  *** &        *** &                  *** &        *** &                  *** \\
            &&  (9.6E-04) &            (5.0E-04) &  (9.6E-04) &            (5.0E-04) &  (9.6E-04) &            (5.0E-04) &  (9.6E-04) &            (5.0E-04) \\
Degree &&     -0.698 &                0.099 &     -0.663 &                0.039 &     -0.372 &                0.031 &     -0.413 &                0.004 \\
            &&        *** &                  *** &        *** &                  *** &        *** &                  *** &        *** &                  *** \\
            &&  (9.6E-04) &            (5.0E-04) &  (9.6E-04) &            (5.0E-04) &  (9.6E-04) &            (5.0E-04) &  (9.6E-04) &            (5.0E-04) \\
Eigenvector &&     -0.868 &                0.110 &     -0.786 &                0.040 &     -0.362 &               -0.016 &     -0.490 &               -0.010 \\
            &&        *** &                  *** &        *** &                  *** &        *** &                  *** &        *** &                  *** \\
            &&  (9.6E-04) &            (5.0E-04) &  (9.6E-04) &            (5.0E-04) &  (9.6E-04) &            (5.0E-04) &  (9.6E-04) &            (5.0E-04) \\
PagerRank &&     -0.777 &                0.094 &     -0.755 &                0.039 &     -0.491 &                0.038 &     -0.535 &                0.006 \\
            &&        *** &                  *** &        *** &                  *** &        *** &                  *** &        *** &                  *** \\
            &&  (9.6E-04) &            (5.0E-04) &  (9.6E-04) &            (5.0E-04) &  (9.6E-04) &            (5.0E-04) &  (9.6E-04) &            (5.0E-04) \\
\hline
\end{tabular}
}

*** $ = p< 0.001$
\end{table}
%

%
\begin{table}
\caption{Results for the model in \refequation{eq:modelFourGraphs}. Standard errors are listed in parentheses.
\newline Intercept: 1.037 *** (1.0E-04), adjusted $R^2$:  $0.873$, p-value:  $<2.2E-16$}
\label{tab:summaryLFRmu}       
%
%

\resizebox{.7\textwidth}{!}{
\begin{tabular}{lrrrrr}
\hline\noalign{\smallskip}

            &  &  $\sqrt{\alpha}$ & $\sqrt{\alpha} \cdot \lambda$ & $\sqrt{\alpha} \cdot \mu$ & $\sqrt{\alpha} \cdot \mu \cdot \lambda$ \\

\hline
Betweenness &&  -0.606 *** &           -0.008 *** &     0.028 *** &               0.021 *** \\
            &&   (6.4E-04) &            (3.4E-04) &     (1.0E-03) &               (5.7E-04) \\
Closeness &&  -0.239 *** &           -0.046 *** &    -0.311 *** &               0.045 *** \\
            &&   (6.4E-04) &            (3.4E-04) &     (1.0E-03) &               (5.7E-04) \\
Degree &&  -0.356 *** &            0.062 *** &    -0.067 *** &              -0.071 *** \\
            &&   (6.4E-04) &            (3.4E-04) &     (1.0E-03) &               (5.7E-04) \\
Eigenvector &&  -0.298 *** &           -0.020 *** &    -0.230 *** &               0.014 *** \\
            &&   (6.4E-04) &            (3.4E-04) &     (1.0E-03) &               (5.7E-04) \\
PagerRank &&  -0.444 *** &            0.065 *** &    -0.117 *** &              -0.070 *** \\
            &&   (6.4E-04) &            (3.4E-04) &     (1.0E-03) &               (5.7E-04) \\
\hline
\end{tabular}
}

*** $ = p< 0.001$
\end{table}
%


\section{Discussion}

Networks are complex, and it is hard to collect network data without missing any nodes or edges. Previous studies have shown that missing nodes can severely affect the reliability of centrality measures. Most  studies focus, however, on cases in which nodes are missing uniformly at random. Yet in a variety of scenarios, it is reasonable to assume that missing nodes may belong to the same community.

In this study, we investigated the reliability of centrality measures when network data is incomplete and the missing nodes are likely to belong to the same community. 
In addition, we introduced a novel approach, called "community bias", which allows researchers to simulate different levels of measurement error.

In our experiments on uniform and scale-free networks, we observed 
that centrality measures are more reliable when missing nodes are likely to belong to the same community compared to those cases in which nodes are missing uniformly at random.
In scale-free networks, the betweenness centrality, however, becomes less reliable with increasing bias. Moreover, in these networks,  centrality measures are also more reliable in networks with stronger community structure. In contrast, we did not observe this effect for uniform networks.

To the knowledge of the author, this is the first study which examines the effect that missing nodes have if their absence depends on the underlying community structure.
A direct comparison to other studies is therefore difficult. 
In contrast to the present study, Niu et al. \cite{Niu2015} found that the biased manipulation of  networks has more severe consequences than a uniformly random manipulation. It is important to note here that the manipulations in \cite{Niu2015} were applied to the edges; nodes were, however, not considered. Our study shows that an increasing bias is associated with higher reliability, which is a novel finding.
If there are legitimate reasons to assume that nodes that have not been observed during the data collection are more likely to belong to the same community, the impact of missing nodes on the reliability of centrality measures might not be as severe as previous studies have suggested.

These findings are encouraging. Although graphs generated by the LFR model share many properties with real-world networks, it would be interesting to see results based on empirical data.
Furthermore, future work may investigate other types of interdependencies between missing nodes, for example, based on node attributes.

\bibliographystyle{myspmpsci}

\end{document}